\documentclass[conference]{IEEEtran}
\IEEEoverridecommandlockouts

\usepackage{amsmath,amssymb,amsfonts}
\usepackage{graphicx}
\usepackage{booktabs}
\usepackage{multirow}
\usepackage{array}
\usepackage{xcolor}
\usepackage{tikz}
\usepackage{cite}
\usepackage{url}
\usepackage{fix-cm}
\usepackage[nolist]{acronym}

\newcolumntype{L}[1]{>{\raggedright\arraybackslash}p{#1}}

\begin{acronym}[BEREC]
\acro{10G-EPON}{10 Gigabit EPON}
\acro{3GPP}{3rd Generation Partnership Project}
\acro{5G}{Fifth Generation}
\acro{6G}{Sixth Generation}
\acro{A-CPI}{A Controller Plane Interface}
\acro{A-RoF}{Analog Radio-over-Fiber}
\acro{ABNO}{Application-Based Network Operations}
\acro{ADC}{Analog-to-Digital Converter}
\acro{AE}{Allocative Efficiency}
\acro{amc}{adaptive modulation and coding}
\acro{AON}{Active Optical Network}
\acro{AI}{Artificial Intelligence}
\acro{ML}{Machine Learning}
\acro{API}{Application Programming Interface}
\acro{AWG}{Arrayed Waveguide Grating}
\acro{B2B}{Business to Business}
\acro{B2C}{Business to Consumer}
\acro{B-RAS}{Broadband Remote Access Servers}
\acro{BaaS}{Blockchain as a Service}
\acro{BB}{Budget Balance}
\acro{BBF}{BroadBand Forum}
\acro{BBU}{Base Band Unit}
\acro{BER}{Bit Error Rate}
\acro{BEREC}{Body of European Regulators for Electronic Communications}
\acro{BF}{beamforming}
\acro{BFT}{Byzentine fault tolerant}
\acro{BGP}{Border Gateway Protocol}
\acro{BMap}{Bandwidth Map}
\acro{BPM}{Business Process Management}
\acro{BS}{Base Station}
\acro{BSC}{Base Station Controller}
\acro{BtB}{Back to Back}
\acro{BW}{Bandwidth}
\acro{C2C}{Consumer to Consumer}
\acro{C-RAN}{Cloud Radio Access Network}
\acro{C-RoFN}{Cloud-based Radio over optical Fiber Network}
\acro{CA}{Certificate Authority}
\acro{CapEx}{Capital Expenditure}
\acro{CDMA}{code division multiple access}
\acro{CDR}{Call Drop Rate}
\acro{CFO}{carrier frequency offset}
\acro{CIR}{Committed Information Rate}
\acro{CO}{Central Office}
\acro{CoMP}{Coordinated Multipoint}
\acro{COP}{Control Orchestration Protocol}
\acro{CORD}{Central Office Rearchitected as a Data Centre}
\acro{CPE}{Customer Premises Equipment}
\acro{CPRI}{Common Public Radio Interface}
\acro{CPU}{Central Processing Unit}
\acro{CMC}{Conflict Mitigation Controller}
\acro{CMS}{Conflict Management System}
\acro{CQI}{channel quality indicator}
\acro{CR}{cognitive radio}
\acro{CS}{Central Station}
\acro{CSI}{channel state information}
\acro{D2D}{Device to Device}
\acro{D-CPI}{D Controller Plane Interface}
\acro{D-RoF}{Digital Radio-over-Fiber}
\acro{DAC}{Digital-to-Analog Converter}
\acro{DAS}{distributed antenna system}
\acro{DAS}{Distributed Antenna Systems}
\acro{DBA}{Dynamic Bandwidth Allocation}
\acro{DBRu}{Dynamic Bandwidth Report upstream}
\acro{DC}{Direct Current}
\acro{DL}{Downlink}
\acro{DLT}{Distributed Ledger Technology}
\acro{DMT}{Discrete Multitone}
\acro{DNS}{Domain Name Server}
\acro{DPDK}{Data Plane Development Kit}
\acro{DSB}{Double Side Band}
\acro{DSL}{Digital Subscriber Line}
\acro{DSLAM}{Digital Subscriber Line Access Multiplexer}
\acro{DSP}{Digital Signal Processing}
\acro{DSS}{Distributed Synchronization Service}
\acro{DU}{Distributed Unit}
\acro{DUE}{D2D user equipment}
\acro{DWDM}{Dense Wave Division Multiplexing}
\acro{E-CORD}{Enterprise CORD}
\acro{EAL}{Environment Abstraction Layer}
\acro{EDF}{Erbium-Doped Fiber}
\acro{EFM}{Ethernet in the First Mile}
\acro{EMBS}{Elastic Mobile Broadband Service}
\acro{EON}{Elastic Optical Network}
\acro{EPC}{Evolved Packet Core}
\acro{EPON}{Ethernet Passive Optical Network}
\acro{eMBB}{enhanced Mobile Broadband}
\acro{ETSI}{European Telecommunications Standards Institute}
\acro{EVM}{Error Vector Magnitude}
\acro{FANS}{Fixed Access Network Sharing}
\acro{FASA}{Flexible Access System Architecture}
\acro{FBMC}{filter bank multi-carrier}
\acro{FBMC}{Filterbank Multicarrier}
\acro{FCC}{Federal Communications Commission}
\acro{FDD}{Frequency Division Duplex}
\acro{FDMA}{frequency-division multiple access}
\acro{FEC}{Forward Error Correction}
\acro{FFR}{Fractional Frequency Reuse}
\acro{FFT}{Fast Fourier Transform}
\acro{FPGA}{Field-Programmable Gate Array}
\acro{FSO}{Free-Space-Optics}
\acro{FSR}{Free Spectral Range}
\acro{FTN}{faster-than-nyquist}
\acro{FTTC}{Fiber-to-the-Curb}
\acro{FTTH}{Fiber-to-the-Home}
\acro{FTTx}{Fiber-to-the-x}
\acro{G.Fast}{Fast Access to Subscriber Terminals}
\acro{GEM}{G-PON encapsulation method}
\acro{GFDM}{Generalized Frequency Division Multiplexing}
\acro{GMPLS}{Generalized Multi-Protocol Label Switching}
\acro{GPON}{Gigabit Passive Optical Network}
\acro{GPP}{General Purpose Processor}
\acro{GPRS}{General Packet Radio Service}
\acro{GSM}{Global System for Mobile Communications}
\acro{GTP}{GPRS Tunneling Protocol}
\acro{GUI}{Graphical User Interface}
\acro{HA}{Hardware Accelerator}
\acro{HARQ}{Hybrid-Automatic Repeat Request}
\acro{HD}{half duplex}
\acro{HetNet}{heterogeneous networks}
\acro{IaaS}{Infrastructure as a Service}
\acro{I-CPI}{I Controller Plane Interface}
\acro{I2RS}{Interface 2 Routing System}
\acro{IA}{Interference Alignment}
\acro{IAM}{Identity and Access Management}
\acro{IBFD}{in-band full duplex}
\acro{IC}{Incentive Compatibility}
\acro{ICIC}{inter-cell interference coordination}
\acro{ICT}{information and communications technology}
\acro{IEEE}{Institute of Electrical and Electronics Engineers}
\acro{IETF}{Internet Engineering Task Force}
\acro{IF}{Intermediate Frequency}
\acro{IFFT}{inverse FFT}
\acro{ITS}{Intelligent Transport Systems}
\acro{InP}{Infrastructure Provider}
\acro{IoT}{Internet of Things}
\acro{IP}{Internet Protocol}
\acro{IQ}{In-phase/Quadrature}
\acro{IR}{Individual Rationality}
\acro{IRC}{Interference Rejection Combining}
\acro{ISI}{inter-symbol interference}
\acro{ISP}{Internet Service Provider}
\acro{BFT}{Byzantine-Fault-Tolerant}
\acro{IV}{Intelligent Vehicle}
\acro{ITU}{International Telecommunication Union}
\acro{KPI}{Key Performance Indicator}
\acro{KVM}{Kernel-based Virtual Machine}
\acro{L2}{Layer-2}
\acro{L3}{Layer-3}
\acro{LAN}{Local Area Network}
\acro{LO}{Local Oscillator}
\acro{LOS}{Line Of Sight}
\acro{LR-PON}{Long Reach PON}
\acro{LSP-DB}{Label Switched Path Database}
\acro{LTE-A}{Long Term Evolution Advanced}
\acro{LTE}{Long Term Evolution}
\acro{M-CORD}{Mobile CORD}
\acro{M-MIMO}{massive MIMO}
\acro{M2M}{machine-to-machine}
\acro{MAC}{Medium Access Control}
\acro{ME}{Merging Engine}
\acro{MIMO}{multiple input multiple output}
\acro{MME}{Mobility Management Entity}
\acro{mmWave}{millimeter wave}
\acro{MNO}{mobile network operator}
\acro{MPLS}{Multiprotocol Label Switching}
\acro{MRC}{Maximum Ratio Combining}
\acro{MSC}{Mobile Switching Centre}
\acro{MSP}{Membership Service Providers}
\acro{MSR}{Multi-Stratum Resources}
\acro{MTC}{machine type communication}
\acro{mMTC}{massive Machine Type Communications}
\acro{MVNO}{Mobile Virtual Network Operator}
\acro{MVNP}{mobile virtual network provider}
\acro{MZI}{Mach-Zehnder Interferometer}
\acro{MZM}{Mach-Zehnder Modulator}
\acro{NE}{Network Element}
\acro{NFV}{Network Function Virtualization}
\acro{NFVaaS}{Network Function Virtualization as a Service}
\acro{NG-PON2}{Next-Generation Passive Optical Network 2}
\acro{NGA}{Next Generation Access}
\acro{NLOS}{None Line Of Sight}
\acro{NMS}{Network Management System}
\acro{NRZ}{Non Return-to-Zero}
\acro{NTT}{Nippon Telegraph and Telephone}
\acro{OBPF}{Optical BandPass Filter}
\acro{OBSAI}{Open Base Station Architecture Initiative}
\acro{ODN}{Optical Distribution Network}
\acro{OFC}{Optical Frequency Comb}
\acro{OFDM}{orthogonal frequency-division multiplexing}
\acro{OFDMA}{orthogonal frequency-division multiple access}
\acro{OLO}{Other Licensed Operator}
\acro{OLT}{Optical Line Terminal}
\acro{ONF}{Open Networking Foundation}
\acro{ONU}{Optical Network Unit}
\acro{OOB}{out-of-band}
\acro{OOK}{On-off Keying}
\acro{OpenCord}{Central Office Re-Architected as a Data Center}
\acro{OpEx}{Operating Expenditure}
\acro{OSI}{Open Systems Interconnection}
\acro{OSS}{Operations Support Systems}
\acro{OTT}{Over-the-Top}
\acro{OXM}{OpenFlow Extensible Match}
\acro{P2MP}{Ethernet over point-to-multipoint }
\acro{P2P}{Point-to-Point }
\acro{PAM}{Pulse Amplitude Modulation}
\acro{PAPR}{peak-to-average power rating}
\acro{PAPR}{Peak-to-Average Power Ratio}
\acro{PBMA}{Priority Based Merging Algorithm}
\acro{PCE}{Path Computation Elements}
\acro{PCEP}{Path Computation Element Protocol}
\acro{PCF}{Photonic Crystal Fiber}
\acro{PD}{Photodiode}
\acro{PDCP}{RD Control Protocol}
\acro{PDF}{probability distribution function}
\acro{PGW}{Packet Gateway}
\acro{PHY}{physical Layer}
\acro{PIR}{Peak Information Rate}
\acro{PMD}{Polarization Division Multiplexing}
\acro{PON}{Passive Optical Network}
\acro{POTS}{Plain Old Telephone Service}
\acro{PoET}{Proof of Elapsed Time}
\acro{PoW}{Proof of Work}
\acro{PoS}{Proof of Stake}
\acro{PTP}{Precision Time Protocol}
\acro{PWM}{Pulse Width Modulation}
\acro{QAM}{Quadrature Amplitude Modulation}
\acro{QoE}{Quality of Experience}
\acro{QoS}{Quality of Service}
\acro{QPSK}{Quadrature Phase Shift Keying}
\acro{R-CORD}{Residential CORD}
\acro{RA}{Resource Allocation}
\acro{RAN}{Radio Access Network}
\acro{RAT}{Radio Access Technology}
\acro{RB}{resource block}
\acro{RFIC}{radio frequency integrated circuit}
\acro{RIC}{RAN Intelligent Controller}
\acro{RLC}{Radio Link Control}
\acro{RN}{Remote Node}
\acro{ROADM}{Reconfigurable Optical Add Drop Multiplexer}
\acro{RoF}{Radio-over-Fiber}
\acro{RRC}{Radio Resource Control}
\acro{RRH}{Remote Radio Head}
\acro{RRPH}{Remote Radio and PHY Head}
\acro{RRU}{Remote Radio Unit}
\acro{RSOA}{Reflective Semiconductor Optical Amplifier}
\acro{RU}{Remote Unit}
\acro{Rx}{receiver}
\acro{SC-FDMA}{single carrier frequency division multiple access}
\acro{SCM}{single carrier modulation}
\acro{SD-RAN}{Software Defined Radio Access Network}
\acro{SDAN}{Software Defined Access Network}
\acro{SDMA}{Semi-Distributed Mobility Anchoring}
\acro{SDN}{Software Defined Networking}
\acro{SDR}{Software Defined Radio}
\acro{SFBD}{Single Fiber Bi-Direction}
\acro{SGW}{Serving Gateway}
\acro{SI}{self-interference}
\acro{SIC}{self-interference cancellation}
\acro{SIMO}{Single Input Multiple Output}
\acro{SLA}{Service Level Agreement}
\acro{SMF}{Single Mode Fiber}
\acro{SNMP}{Simple Network Management Protocol}
\acro{SNR}{Signal-to-Noise Ratio}
\acro{SP}{Service Providers}
\acro{Split-PHY}{Split Physical Layer}
\acro{SRI-OV}{Single Root Input/Output Virtualization}
\acro{SU}{secondary user}
\acro{SUT}{System Under Test}
\acro{T-CONT}{Transmission Container}
\acro{TC}{Transmission Convergence}
\acro{TCO}{Total Cost of Ownership}
\acro{TD-LTE}{Time Division LTE}
\acro{TDD}{Time Division Duplex}
\acro{TDM}{Time Division Multiplexing}
\acro{TDMA}{Time Division Multiple Access}
\acro{TED}{Traffic Engineering Database}
\acro{TEID}{Tunnel endpoint identifier}
\acro{TLS}{Transport Layer Security}
\acro{TPS}{Transactions per Second}
\acro{TTI}{transmission time interval}
\acro{TWDM}{Time and Wavelength Division Multiplexing}
\acro{Tx}{transmitter}
\acro{UD-CRAN}{Ultra-Dense Cloud Radio Access Network}
\acro{UDP}{User Datagram Protocol}
\acro{UE}{User Equipment}
\acro{UF-OFDM}{Universally Filtered OFDM}
\acro{UFMC}{Universally Filtered Multicarrier}
\acro{UL}{Uplink}
\acro{UMTS}{Universal Mobile Telecommunications System}
\acro{URLLC}{Ultra-Reliable Low-Latency Communication}
\acro{USRP}{Universal Software Radio Peripheral}
\acro{V2I}{Vehicle to Infrastructure}
\acro{V2V}{Vehicle to Vehicle}
\acro{vBBU}{virtualized BBU}
\acro{vBMap}{Virtual Bandwidth Map}
\acro{vBS}{virtual Base station}
\acro{VCG}{Vickery-Clarke-Groves}
\acro{vCPE}{virtual CPE}
\acro{vCPU}{Virtual Central Processing Unit}
\acro{vDBA}{virtual DBA}
\acro{VDSL2}{Very-high-bit-rate digital subscriber line 2}
\acro{VLAN}{Virtual Local Area Network}
\acro{VM}{Virtual Machine}
\acro{VNF}{Virtual Network Function}
\acro{VNI}{visual networking index}
\acro{VNO}{Virtual Network Operator}
\acro{VNTM}{Virtual Network Topology Manager}
\acro{vOLT}{virtual OLT}
\acro{VPE}{virtual Provider Edge}
\acro{VPN}{Virtual Private Network}
\acro{V2X}{Vehicle to Everything}
\acro{VTN}{Virtual Tenant Network}
\acro{VULA}{Virtual Unblunded Local Access}
\acro{WAN}{Wide Area Network}
\acro{WAP}{Wireless Access Point}
\acro{WCDMA}{wide-band code division multiple access}
\acro{WDM-PON}{Wavelength Division Multiplexing - Passive Optical Network}
\acro{WDM}{Wavelength Division Multiplexing}
\acro{WiMAX}{Worldwide Interoperability for Microwave Access}
\acro{WRPR}{Wired-to-RF Power Ratio}
\acro{XG-PON}{10 Gigabit PON}
\acro{XGEM}{XG-PON encapsulation method}
\acro{XOS}{XaaS Operating System}

\acro{RIC}{RAN Intelligent Controller}
\acro{Near-RT RIC}{Near Real Time RAN Intelligent Controller}
\acro{Non-RT RIC}{Non Real Time RAN Intelligent Controller}
\acro{xApp}[xApp]{RAN control application}
\acro{rApp}{Remote Application}
\acro{dApp}{Distributed Application}
\acro{CMS}{Conflict Management System}
\acro{EG}{Eisenberg--Gale}
\acro{NSWF}{Nash Social Welfare Function}
\acro{CMC}{Conflict Mitigation Controller}
\acro{CDC}{Conflict Detection Controller}
\acro{PMon}{Performance Monitoring}
\acro{RCP}{Recently Changed Parameter}
\acro{PGD}{Parameter Group Definition}
\acro{RCPG}{Recently Changed Parameter Group}
\acro{DCKD}{Decision Correlated with KPI Degradation} 
\acro{KDO}{KPI Degradation Occurrences}
\acro{MNO}{Mobile Network Operator}
\acro{SDL}{Shared Data Layer}
\acro{SON}{Self Organising Network}
\acro{SF}{SON Function}
\acro{AM}{Arithmetic Mean}
\acro{PKR}{Parameter and KPI Ranges}
\acro{KPI}{Key Performance Indicator}
\acro{ICP}{Input Control Parameter}
\acro{RF}{Radio Frequency}
\acro{C-RAN}{Cloud Radio Access Network}
\acro{RL}{Reinforcement Learning}
\acro{MARL}{Multi Agent Reinforcement Learning}
\acro{CCO}{Capacity and Coverage Optimization}
\acro{ES}{Energy Saving}
\acro{MRO}{Mobility Robustness Optimization}
\acro{MLB}{Mobility Load Balancing}
\acro{V-RAN}{Virtualized Radio Access Network}
\acro{O-RAN}{Open Radio Access Network}
\acro{D-RAN}{Distributed Radio Access Network}
\acro{5GB}{5G and Beyond}
\acro{O-RU}{Open Radio Unit}
\acro{O-DU}{Open Distribution Unit}
\acro{O-CU}{Open Centralised Unit}
\acro{O-CU-CP}{Open Centralised Unit - Control Plane}
\acro{O-CU-UP}{Open Centralised Unit - User Plane}
\acro{D-SON}{Distributed SON}
\acro{C-SON}{Centralised SON}
\acro{CIO}{Cell Individual Offset}
\acro{CBR}{Call Block Ratio}
\acro{HOR}{Handover Ratio}
\acro{CAN}{Cognitive Autonomous Network}
\acro{CF}{Cognitive Function}
\acro{QACM}{QoS-Aware Conflict Mitigation}
\acro{ANN}{Artificial Neural Network}
\acro{PR}{Polynomial Regression}
\acro{MSE}{Mean Squared Error}
\acro{EVS}{Explained Variance Score}
\acro{CS}{Conflict Supervision}
\acro{TXP}{Tx Power}
\acro{TTT}{Time-To-Trigger}
\acro{CIO}{Cell Individual Offset}
\acro{NL}{Neighbour List}
\acro{RET}{Radio Electrical Tilt}
\acro{HYS}{Handover Hysteresis}
\acro{QACMP}{QACM Priority}
\acro{DRL}{Deep Reinforcement Learning}
\acro{gNB}{Next-generation NodeB}
\acro{RSRP}{Reference Signal Received Power}
\acro{P2K}{parameter-to-KPI}
\acro{P2X}{parameter-to-xApp}
\acro{K2X}{KPI-to-xApp}
\acro{VK}{violating-KPI}
\acro{UP}{Uplink}
\acro{DL}{Downlink}
\acro{AIMLFW}{Open RAN AI/ML framework}
\acro{KPM}{Key Performance Measurement}
\acro{E2SM}{E2 Service Model}
\acro{EI}{Enrichment Information}
\acro{NWDAF}{Network Data Analytics Function}
\acro{PCF}{Policy Control Function}
\acro{PPO}{Proximal Policy Optimization}
\acro{GPU}{Graphics Processing Unit}
\acro{SIMD}{Single Instruction, Multiple Data}
\acro{GNN}{Graph Neural Network}
\acro{TL}{Traffic Load}
\acro{SMOTE}{Synthetic Minority Over-sampling}
\acro{ECDF}{Empirical Cumulative Distribution Function}
\acro{OpenRAN}[Open RAN]{Open Radio Access Network}
\acro{CU}{Centralised Unit}
\acro{DDQN}{Double Deep Q-Network}
\acro{DQN}{Deep Q-Network}
\acro{BFP}{Block Floating Point}
\acro{FR1}{Frequency Range 1}
\acro{CDF}{Cumulative Distribution Function}
\acro{KS}{Kolmogorov--Smirnov}
\acro{INT}{integrated distance}
\acro{SINR}{Signal-to-Interference-plus-Noise Ratio}
\acro{HOF}{Handover Failure}
\acro{CI}{Confidence Interval}
\acro{SD}{Standard Deviation}
\acro{GT}{game-theoretic}
\acro{FC}{fully connected}
\end{acronym}

\begin{document}

\title{%
\begin{tikzpicture}[remember picture, overlay]
    \node[anchor=north, yshift=-0.5cm] at (current page.north) {\fbox{\parbox{\textwidth}{\centering\small\color{red}This version of the paper has been accepted for presentation at IEEE WCSP 2026. It is subject to IEEE copyright and may be removed upon request.}}};
\end{tikzpicture}%
Cost-Aware Conflict Triage for xApp Control Conflicts in Open RAN \\

\thanks{Funding for this research is provided by the European Union's Horizon Europe research and innovation program through the Marie Sklodowska-Curie SE grant under the agreement RE-ROUTE No 101086343.}
}

\author{\IEEEauthorblockN{Abdul Wadud\IEEEauthorrefmark{1}\IEEEauthorrefmark{2},
Fatemeh Golpayegani\IEEEauthorrefmark{1}, and Nima Afraz\IEEEauthorrefmark{1}}
\IEEEauthorblockA{\IEEEauthorrefmark{1}School of Computer Science, University College Dublin, Dublin, Ireland}
\IEEEauthorblockA{\IEEEauthorrefmark{2}Bangladesh Institute of Governance and Management, Dhaka, Bangladesh}
\IEEEauthorblockA{abdul.wadud@ucdconnect.ie, \{fatemeh.golpayegani, nima.afraz\}@ucd.ie}}

\maketitle

\begin{abstract}
Open Radio Access Network (Open RAN) enables multi-vendor deployments where concurrent \acp{xApp} independently optimise mobility, load, coverage, and energy within the \ac{Near-RT RIC}. When \acp{xApp} act on overlapping control parameters, their isolated loops produce conflicting decisions that degrade the network \acp{KPI}. Existing conflict-evaluation methods estimate severity but do not answer whether the intervention is cost-justified. This paper proposes a cost-aware conflict triage framework that pairs ECDF-based severity evaluation with an event-level decision layer. The framework groups detected conflict records into episodes, extracts nine features, estimates per-mechanism costs and benefits, and assigns one of three actions: tolerate, defer, or mitigate. Evaluated in a simulator-driven Open RAN setting over 30 Monte Carlo runs with four concurrent \acp{xApp}, the framework defers 66.8\,\%, mitigates 19.8\,\%, and tolerates 13.4\,\% of detected episodes, achieving a mean net utility of 0.132 per intervention, 3.9$\times$ higher than an always-mitigate baseline, by avoiding low-value interventions through cost-aware selectivity.
\end{abstract}


\acresetall 

\section{Introduction}
\label{sec:intro}
\IEEEPARstart{O}{pen} \ac{RAN} makes the radio access network more flexible by allowing operators to deploy multiple control applications from different vendors, called \acp{xApp}, inside the \ac{Near-RT RIC}. Different \acp{xApp} can independently optimise distinct aspects of the network, such as mobility robustness, energy saving, load balancing, or \ac{QoS}. However, this flexibility also creates a new challenge. Two or more \acp{xApp} may attempt to influence the same control parameter, the same radio resource, or the same \ac{KPI} simultaneously. When that happens, conflicts arise, which can lead to network-wide performance degradation. 

Prior work covers conflict detection and taxonomy \cite{wadud2024xapp}, rule-based and policy-based mitigation \cite{adamczyk2023conflict, del2024pacifista}, optimization-based and learning-based resolution \cite{wadud2024qacm, wadud2023conflict}, and proactive avoidance via scheduling or digital twins \cite{cinemre2025xapp, giannopoulos2025comix}. However, most treat mitigation as automatic once a conflict is flagged. Notably, PACIFISTA \cite{del2024pacifista} provides robust severity evaluation but does not assess cost-justification, and QACM \cite{wadud2024qacm} selects the best mechanism but assumes mitigation is always warranted. The proposed triage layer fills this gap by adding the cost-awareness step that neither addresses.

Mitigation carries real operational cost: processing overhead, signalling load, added latency, and disruption to ongoing control loops \cite{slimani2006conflict, katsigiannis2015cost}. Conflicts that produce only transient or minor degradation may not justify this cost, making \emph{whether} to mitigate a distinct and necessary question alongside \emph{how} to mitigate. Motivated by these observations, we introduce \emph{conflict triage} as a decision layer for \ac{xApp} conflict management. Once a conflict is detected, the controller chooses to \texttt{tolerate} (take no corrective action), \texttt{defer} (postpone mitigation and re-observe), or \texttt{mitigate} (intervene immediately). The key idea is simple: mitigate only when the expected benefit exceeds the expected cost. Built on top of PACIFISTA-style \ac{ECDF} severity evaluation \cite{del2024pacifista}, the proposed framework extracts nine episode-level features from detected conflict records, estimates the expected loss of inaction, compares it against per-mechanism cost and benefit across six candidates, and selects the action with the highest expected utility.

The main contributions of this paper are:
\begin{itemize}
    \item We introduce conflict triage as a cost-aware decision layer for \ac{xApp} conflict handling in Open RAN.
    \item We combine an upstream PACIFISTA-style severity prior with an event-level utility model that supports tolerate, defer, and mitigate decisions.
    \item We define a structured mitigation cost model that includes computation, signalling, latency, stability, and opportunity cost.
    \item We demonstrate via 30-run Monte Carlo evaluation that the triage layer escalates high-severity, high-degradation episodes while deferring borderline cases, confirming internal consistency.
\end{itemize}

\section{System Model and Problem Formulation}
\label{sec:system_model}

We consider an Open \ac{RAN} deployment comprising one central unit (CU), multiple distributed units (DUs), and radio units (RUs), all managed through the \ac{Near-RT RIC}. The network is observed at discrete time steps of duration $\Delta_t$, during which each \ac{xApp} may issue at most one control action. Multiple \acp{xApp} operate simultaneously on shared radio-control parameters through the \ac{Near-RT RIC}. Let $\mathcal{X}=\{\text{MRO},\text{MLB},\text{CCO},\text{ES}\}$ denote the set of \acp{xApp} considered in this work: \ac{MRO} optimises handover parameters to reduce radio link failures; \ac{MLB} redistributes load across cells by adjusting handover offsets; \ac{CCO} tunes transmit power for coverage and capacity trade-offs; and \ac{ES} deactivates underloaded cells to reduce energy consumption. Each \ac{xApp} issues control actions according to its own objective, while the resulting network behaviour is observed through \ac{KPI} traces. Because the \acp{xApp} do not optimise the same objective, an action taken by one \ac{xApp} may degrade the \ac{KPI} targeted by another \ac{xApp}. This creates an \ac{xApp} conflict.

The system observes two main inputs over time: \ac{xApp} action logs and \ac{KPI} traces. From these inputs, the upstream monitoring layer identifies conflict evidence and associates it with a logical conflict event. We model one conflict event as an episode, $e=(t_e,x_i,x_j,p_e,k_e,c_e)$, where $t_e$ is the event time, $x_i$ is the instructing \ac{xApp}, $x_j$ is the affected \ac{xApp}, $p_e$ is the changed control parameter, $k_e$ is the degraded \ac{KPI}, and $c_e \in \{\text{direct},\text{indirect},\text{implicit}\}$ is the conflict type.

For each episode, the controller builds a feature vector, $\phi_e=[s_e,\delta_e,\rho_e,\omega_e,\kappa_e,\lambda_e,u_e,r_b(e),r_h(e)]$, where $s_e \in [0,1]$ is the PACIFISTA-derived severity prior for the instructing--affected \ac{xApp} pair; $\delta_e \in [0,1]$ is the normalised \ac{KPI} degradation magnitude, computed as the mean relative \ac{KPI} drop across all affected \acp{KPI} during the conflict window; $\rho_e \in [0,1]$ is the persistence score, reflecting the fraction of the observation window over which the degradation is sustained; $\omega_e \in [0,1]$ is the impact scope, capturing the proportion of affected cells or slices; $\kappa_e \in [0,1]$ is the operator-assigned \ac{KPI} criticality weight; $\lambda_e \in [0,1]$ is the normalised system control-load state at the time of the episode; $u_e \in [0,1]$ is the uncertainty estimate derived from the variance of the severity prior and \ac{KPI} measurement noise; and $r_b(e)$ and $r_h(e)$ are the benign and harmful recurrence counts, respectively. The first seven features are normalised to $[0,1]$; recurrence is represented by counts and bounded through the recurrence factor below. This design allows the controller to distinguish between conflicts that repeatedly self-resolve and conflicts that repeatedly cause damage.

The controller can respond to an episode in three ways: tolerate it, defer the decision, or mitigate it. If mitigation is selected, one mechanism is chosen from the set of mitigation methods $\mathcal{M}$. In this paper, \[\mathcal{M}=\{\texttt{cancel},\texttt{dampen}, \texttt{priority}, \texttt{nswf}, \texttt{eg}, \texttt{qacm}\}.\]

The first three are unilateral action-modification mechanisms, while the last three are game-theoretic compromise mechanisms. \texttt{cancel} suppresses the conflicting action entirely when the safest choice is to block it. \texttt{dampen} reduces the strength of the action so that its negative impact is softened rather than fully removed. \texttt{priority} replaces the proposed action with a preferred alternative according to the controller's resolution logic \cite{adamczyk2023conflict}. \texttt{nswf} applies a Nash social welfare objective to favour balanced outcomes across competing \acp{xApp} \cite{wadud2023conflict}. \texttt{eg} uses an egalitarian bargaining rule to promote fairness between conflicting objectives \cite{wadud2023conflict}. \texttt{qacm} provides a compromise-oriented mechanism that explicitly trades off benefit and cost under conflict \cite{wadud2024qacm}. These six mechanisms were chosen to cover both simple, low-overhead action modifications and cooperative, utility-driven compromises. Table~\ref{tab:cost_coeffs} and Fig.~\ref{fig:mechanism_winrate} use the implementation labels \texttt{cancel\_latest}, \texttt{dampen\_latest}, and \texttt{priority\_override} (shortened to \texttt{override} in Table~\ref{tab:cost_coeffs}) for \texttt{cancel}, \texttt{dampen}, and \texttt{priority}, respectively. The opportunity cost component $C_\text{opp}(e,m)$ in Eq.~\ref{eq:cmit} captures the foregone benefit of not applying an alternative mechanism, i.e., the utility difference between $m$ and the next-best candidate. If a mechanism has no addressable action for a given $c_e$ (e.g., \texttt{cancel} applied to a conflict where no single action can be suppressed), its $\eta_{m,c_e}$ is set to zero, removing its loss-reduction benefit; its score still includes uncertainty and cost penalties.

To decide whether intervention is worthwhile, we model the expected unmanaged loss, mitigation cost, residual loss after mitigation, mitigation benefit, and net utility as follows. The coefficients in these models can either be set by the \ac{MNO} according to operational priorities or calibrated offline from data. In this paper, we calibrate coefficient values using an offline multi-objective random-search procedure with Pareto filtering and weighted compromise selection.
\begin{equation}
\begin{aligned}
L_{\mathrm{conf}}(e)
&= s_e \delta_e
\left( 1+\beta_{\rho}\rho_e+\beta_{\omega}\omega_e \right. \\
&\qquad \left. +\beta_{\kappa}\kappa_e+\beta_{\lambda}\lambda_e \right)
\left( 1+\beta_h r_h(e)-\beta_b r_b(e) \right)
\end{aligned}
\label{eq:lconf}
\end{equation}

\begin{equation}
\begin{aligned}
C_{\mathrm{mit}}(e,m)
&= C_{\mathrm{comp}}(e,m)+C_{\mathrm{sig}}(e,m) \\
&\quad + C_{\mathrm{lat}}(e,m)+C_{\mathrm{stab}}(e,m) \\
&\quad + C_{\mathrm{opp}}(e,m)
\end{aligned}
\label{eq:cmit}
\end{equation}

\begin{equation}
L_{\mathrm{res}}(e,m)
=
\bigl(1-\eta_{m,c_e}\bigr)L_{\mathrm{conf}}(e)+\xi_m u_e
\label{eq:lres}
\end{equation}

\begin{equation}
B(e,m)=L_{\mathrm{conf}}(e)-L_{\mathrm{res}}(e,m)
\label{eq:benefit}
\end{equation}

\begin{equation}
U(e,m)=B(e,m)-C_{\mathrm{mit}}(e,m)
\label{eq:utility}
\end{equation}

Here, $L_{\mathrm{conf}}(e)$ represents the expected unmanaged loss, the cumulative \ac{KPI} degradation that would accrue if the conflict is left unmitigated. $C_{\mathrm{mit}}(e,m)$ is the operational cost of applying mechanism $m$, comprising computation, signalling, latency, stability, and opportunity cost components (Eq.~\ref{eq:cmit}), reflecting the cost dimensions most consistently cited for real-time RIC control actions \cite{slimani2006conflict,katsigiannis2015cost} rather than an exhaustive taxonomy. $L_{\mathrm{res}}(e,m)$ is the remaining unmanaged loss after mitigation, $B(e,m)$ is the mitigation benefit, and $U(e,m)$ is the final net utility. The coefficients $\beta_{\rho}$, $\beta_{\omega}$, $\beta_{\kappa}$, and $\beta_{\lambda}$ control the relative influence of persistence, scope, \ac{KPI} criticality, and load state, respectively, while $\beta_h$ and $\beta_b$ control the effect of harmful and benign recurrence. All $\beta$ coefficients are non-negative (see Table~\ref{tab:cost_coeffs} in Section~\ref{sec:experiments} for the calibrated values), and the recurrence factor $(1 + \beta_h r_h(e) - \beta_b r_b(e))$ is clipped to $[0, 2]$ to prevent negative loss estimates under high benign recurrence. The effectiveness parameter $\eta_{m,c_e} \in [0,1]$ captures the expected fraction of unmanaged loss eliminated by mechanism $m$ for conflict type $c_e$; it is calibrated offline for each mechanism--conflict-type pair using historical simulation data. The parameter $\xi_m \geq 0$ is a mechanism-specific uncertainty penalty reflecting the reliability of each mechanism's effect estimate (calibrated values in Table~\ref{tab:cost_coeffs}). Thus, larger $\eta_{m,c_e}$ reduces the residual loss, while larger $\xi_m$ makes the controller more cautious under uncertainty.

The loss model is an operator-defined surrogate, not a physical law. The product $s_e\delta_e$ requires both prior conflict evidence and observed degradation; non-negative linear weights make contextual effects monotone and interpretable. The clipped recurrence factor bounds history effects. This structure supports transparent calibration but does not establish that the chosen form predicts actual network loss better than alternative models.

The problem is therefore to assign a triage action to each conflict episode such that costly intervention is avoided when it is unnecessary, while harmful conflicts are still handled in time. Formally, the controller first selects the best mitigation candidate as $m^*(e)=\arg\max_{m\in\mathcal{M}} U(e,m)$, and then maps the resulting utility to one of the three triage actions $\alpha(e)\in\{\texttt{tolerate},\texttt{defer},\texttt{mitigate}\}$. The operational mapping from utility and uncertainty to the final triage action is defined in the proposed framework in Section~\ref{sec:triage_framework}.

\begin{figure}[t]
    \centering
    \includegraphics[scale=0.45]{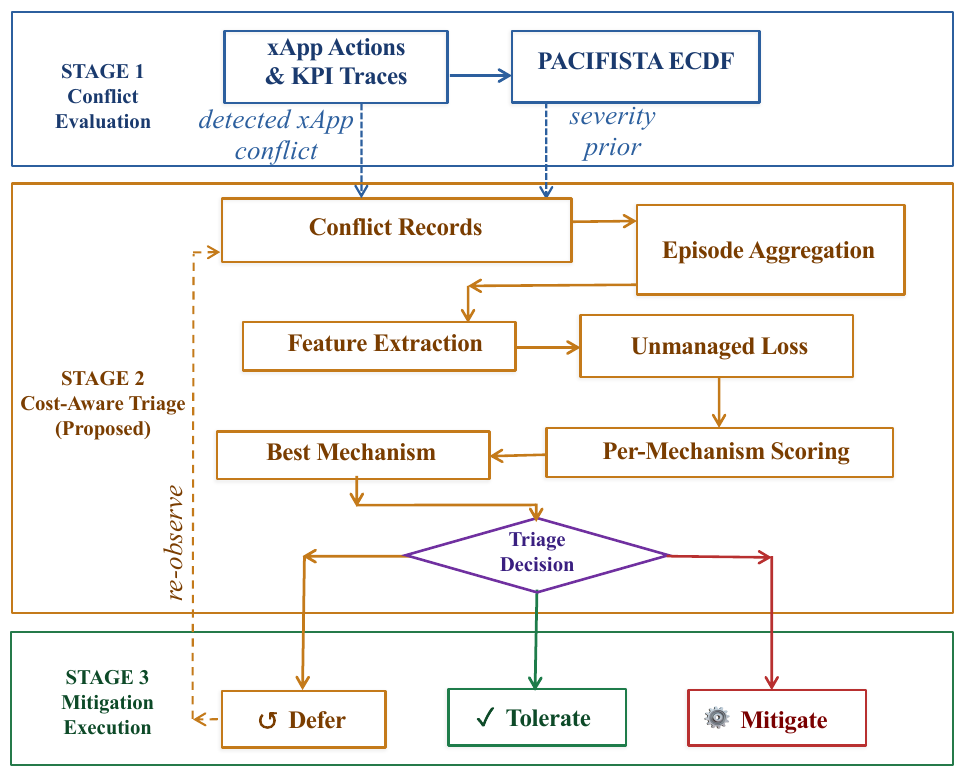}
    \vspace{-0.2cm}
    \caption{Proposed cost-aware conflict triage pipeline.}
    \label{fig:triage_pipeline}
\end{figure}

\section{Proposed Cost-Aware Conflict Triage Framework}
\label{sec:triage_framework}

Fig.~\ref{fig:triage_pipeline} shows the proposed pipeline. The workflow has three stages: \emph{conflict evaluation}, \emph{cost-aware triage}, and \emph{mitigation execution}. The first stage is upstream and provides conflict evidence. The second stage is the main contribution: rather than mitigating every detected conflict, the controller compares expected benefit with expected cost and chooses among tolerate, defer, and mitigate. The third stage applies the final decision through the conflict management component.

\subsection{Stage 1: Upstream Conflict Evaluation}

Action logs and \ac{KPI} traces feed the upstream detector and PACIFISTA-style \ac{ECDF} profiling. Profiling supplies a pairwise severity prior $s_e$; detection supplies raw conflict records. Neither output alone triggers intervention: high severity with small observed degradation can still yield low utility in Eqs.~\ref{eq:lconf}--\ref{eq:utility}.

\subsection{Stage 2: Cost-Aware Triage}
Records sharing an instructing--affected \ac{xApp} pair within the aggregation window $\Delta_t$ are grouped into an episode. The controller extracts $\phi_e$, computes unmanaged loss once using Eq.~\ref{eq:lconf}, and scores each candidate using Eqs.~\ref{eq:cmit}--\ref{eq:utility}. It retains the best mechanism $m^*(e)$ and its utility $U^*(e)=U(e,m^*(e))$ for the decision stage. Benign and harmful recurrence enter the loss estimate separately, distinguishing self-resolving events from repeated damage.

\subsection{Stage 3: Triage Decision and Mitigation Execution}
Let $u_e$ denote the uncertainty of episode $e$. The thresholds $\tau_l$ and $\tau_h$ are the lower and upper utility thresholds, respectively, and $u_h$ is the uncertainty threshold. Thus, $\tau_l$ separates clearly non-beneficial cases from borderline ones, $\tau_h$ separates borderline cases from clearly beneficial ones, and $u_h$ limits mitigation under high uncertainty. The final triage action is then defined as
{\small
\begin{equation}
\alpha(e)=
\begin{cases}
\texttt{mitigate with } m^*(e), &
U^*(e)>\tau_h \;\land\; u_e\le u_h,\\[2pt]
\texttt{defer}, &
\tau_l \le U^*(e)\le \tau_h,\\[2pt]
\texttt{defer}, &
U^*(e)>\tau_h \;\land\; u_e>u_h,\\[2pt]
\texttt{tolerate}, &
U^*(e)<\tau_l.
\end{cases}
\label{eq:triage_rule}
\end{equation}
}

Deferral is useful, but it should not continue forever. Therefore, the framework uses bounded deferral. If an episode is deferred repeatedly and its deferral count reaches the maximum allowed limit $D_{\max}$, the controller escalates the event and applies the best available mitigation. This keeps the system from ignoring persistent, harmful conflicts simply because early evidence was uncertain. The thresholds $\tau_l$, $\tau_h$, $u_h$, and $D_{\max}$ are calibrated alongside the model coefficients $\beta$ using the offline random-search procedure described in Section~\ref{sec:system_model}. Note that $\tau_l$ is set to a small negative value ($-0.02$), so that only episodes with clearly negative expected utility are assigned \texttt{tolerate}; episodes with slightly negative or borderline utility in $[\tau_l, 0]$ are deferred for re-observation rather than immediately dismissed.

\subsection{Computational Cost}
With episode features and calibrated coefficients available, loss estimation takes constant time. Scoring and selecting among $M=|\mathcal{M}|$ mechanisms takes $O(M)$ time and can use $O(1)$ working storage by retaining the best score. Here $M=6$; scoring uses arithmetic and comparisons without online training. This bound excludes upstream detection, trace aggregation, and execution of the selected mechanism, including any optimisation solver. End-to-end inference latency is not measured; the simulation's observation interval does not establish a deployment latency guarantee.

\section{Experimental Methodology}
\label{sec:experiments}

This section evaluates the proposed cost-aware conflict triage framework in a simulator-driven Open \ac{RAN} setting. We use MATLAB 5G Toolbox to represent an O-RAN 7.2x architecture \cite{wadud2026ai2}. The evaluation assesses the internal behaviour of the policy layer; it does not constitute validation of a deployed Near-RT RIC or live E2 control loop.

\subsection{Simulation Setting}
For the experimental setup, we consider a four-xApp scenario presented in \cite{wadud2024qacm} that incurs direct and indirect conflict due to their \ac{xApp}-parameter-\ac{KPI} relationship dynamics. These four interacting \acp{xApp} are: \ac{MRO}, \ac{MLB}, \ac{CCO}, and \ac{ES}. These \acp{xApp} operate on shared radio-control parameters and generate action logs together with \ac{KPI} traces, which form the inputs to the upstream conflict-evaluation stage and the proposed triage layer.

For the evaluation, we use a fixed baseline network scenario with one CU, three DUs, and nine RUs under a 7.2x split. The radio profile is FR1 at 3.5~GHz with 100~MHz bandwidth and simplified fixed beamforming. Each run serves 100 users with mixed mobility patterns. The controller implementations are: DDQN-based MRO \cite{wang2024double}, supermodular load balancing for MLB \cite{zafar2025conflict}, SVD-based power recommendation for CCO \cite{hassan2024machine}, and VBP-style energy saving for ES \cite{zafar2025conflict}. The DDQN-MRO agent completes 12{,}461 gradient updates ($\varepsilon_{\text{final}}=0.050$), after which the trained policy is used in the evaluation. Table~\ref{tab:triage_thresholds} lists the triage threshold and deferral parameters used in all experiments. These values were selected by the offline calibration procedure (24 candidates, 3 MC runs each, single Pareto-front winner) and kept fixed across all 30 evaluation runs. Table~\ref{tab:cost_coeffs} lists the calibrated loss-weight and uncertainty-penalty coefficients. Calibration and evaluation use the same fixed topology and operating profile.

\begin{table}[t]
\centering
\caption{Triage threshold and deferral parameters.}
\label{tab:triage_thresholds}
\setlength{\tabcolsep}{4pt}
\begin{tabular}{lll}
\hline
Parameter & Value & Role \\
\hline
$\tau_l$ (lower threshold) & $-0.02$ & Tolerate if $U^*(e)<\tau_l$ \\
$\tau_h$ (upper threshold) & $0.18$  & Mitigate if $U^*(e)>\tau_h$, $u_e\le u_h$ \\
$u_h$ (uncertainty gate)   & $0.70$  & Defer if $u_e>u_h$ \\
$D_{\max}$ (max deferrals) & $2$     & Escalate after 2 deferrals ($20\,ms$) \\
\hline
\end{tabular}
\end{table}

\begin{table}[t]
\centering
\caption{Loss-weight ($\beta$) and uncertainty-penalty ($\xi_m$) coefficients.}
\label{tab:cost_coeffs}
\setlength{\tabcolsep}{4pt}
\begin{tabular}{llll}
\hline
Coeff. & Value & Coeff. & Value \\
\hline
$\beta_{\rho}$ (persistence) & 0.35 & $\xi_{\texttt{cancel}}$ & 0.10 \\
$\beta_{\omega}$ (scope)      & 0.25 & $\xi_{\texttt{dampen}}$ & 0.06 \\
$\beta_{\kappa}$ (KPI crit.) & 0.40 & $\xi_{\texttt{override}}$ & 0.08 \\
$\beta_{\lambda}$ (load)      & 0.25 & $\xi_{\texttt{nswf}}$   & 0.08 \\
$\beta_h$ (harmful recur.)    & 0.18 & $\xi_{\texttt{eg}}$     & 0.07 \\
$\beta_b$ (benign recur.)     & 0.12 & $\xi_{\texttt{qacm}}$   & 0.05 \\
\hline
\end{tabular}
\end{table}

\subsection{Monte Carlo Protocol}
To evaluate stability, we run the full pipeline over 30 Monte Carlo trials with different random seeds and report the mean, standard deviation, and 95\,\% confidence interval for each metric. Unless otherwise stated, the triage layer uses the calibrated coefficient setting stated in Section~\ref{sec:triage_framework}.

\begin{figure}[t]
    \centering
    \includegraphics[scale=0.4]{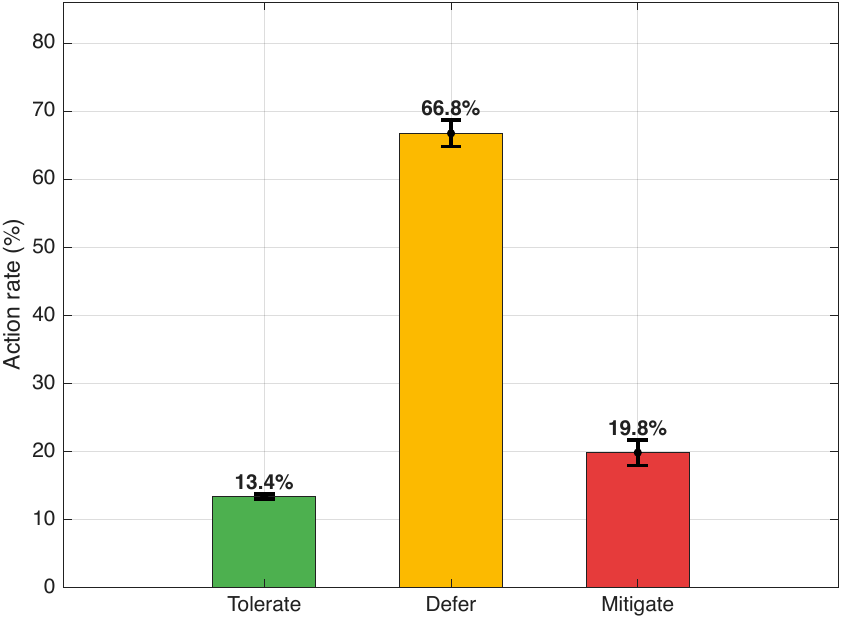}
    \vspace{-0.3cm}
    \caption{Triage action distribution. \texttt{Defer} dominates at 66.8\,\%, confirming selective operation: the majority of detected conflicts are held for re-observation rather than immediately mitigated.}
    \vspace{-0.3cm}
    \label{fig:mc_summary_overview}
\end{figure}

\section{Results and Discussion}
\label{sec:results}

\subsection{Overall Triage Behaviour and Stability}

The first result is that the proposed framework behaves as a selective decision layer rather than an always-on mitigation trigger. Across 30 Monte Carlo runs, the controller detects on average 118.3 conflict episodes per run (std 33.2, 95\,\%~CI [105.9,130.7]). On average, \texttt{defer} is the dominant action at 66.8\,\% (CI [64.8,\,68.7]), followed by \texttt{mitigate} at 19.8\,\% (CI [17.9,\,21.7]) and \texttt{tolerate} at 13.4\,\% (CI [13.0,\,13.8]), as shown in Fig.~\ref{fig:mc_summary_overview}. Translating the rates to absolute counts: the framework defers a mean of 79.0 episodes, mitigates 23.4, and tolerates 15.8 per run.

These 118.3 episodes arise from six structurally conflicting \ac{xApp} pairs across all four \acp{xApp}. MRO-facing pairs show the highest PACIFISTA severity priors (up to 0.241 for MRO--MLB) because MRO shares \ac{TXP} and handover-offset \acp{ICP} with every other \ac{xApp}; the lowest severity pair (MLB--ES) scores 0.018.

This pattern is consistent with the framework's design goal of reserving intervention for episodes where the cost is justified. Many detected conflicts are neither clearly harmless nor clearly severe enough to justify immediate action. Deferral provides a useful intermediate state that allows the controller to accumulate evidence rather than spending mitigation cost prematurely, while bounded deferral ($D_{\max}=2$) prevents indefinite inaction. The confidence intervals in Fig.~\ref{fig:mc_summary_overview} describe variation across seeds under the fixed scenario; they do not establish robustness across operating conditions.

\subsection{Utility of Cost-Aware Triage}

\begin{figure}[t]
    \centering
    \includegraphics[scale=0.5]{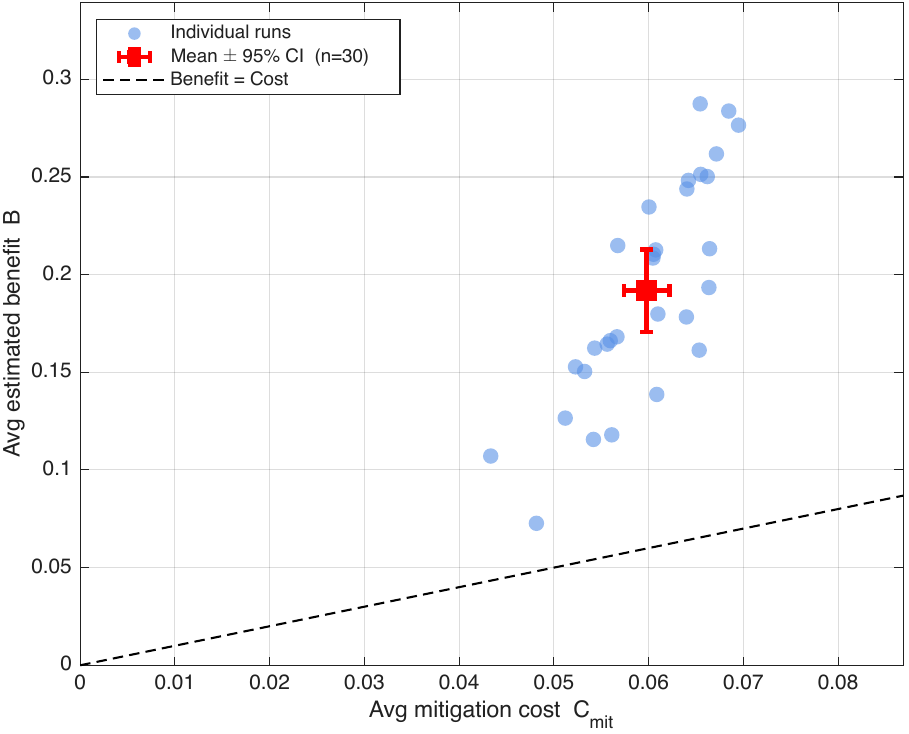}
    \vspace{-0.3cm}
    \caption{Average estimated benefit vs.\ mitigation cost per run. All runs sit above the benefit-equals-cost line; the mean operating point (red square) confirms positive net utility across runs.}
    \label{fig:benefit_cost}
\end{figure}

The second result is that selected interventions remain beneficial on average. Over 30 runs, the mean mitigation benefit is 0.192 (CI [0.171,\,0.213]), the mean residual loss is 0.076 (CI [0.069,\,0.083]), and the mean net utility per mitigated episode is 0.132 (CI [0.113,\,0.151]). The benefit-versus-cost view in Fig.~\ref{fig:benefit_cost} shows that the mean operating point lies clearly above the benefit-equals-cost line, confirming worthwhile interventions on average.

\begin{figure}[t]
    \centering
    \includegraphics[scale=0.35]{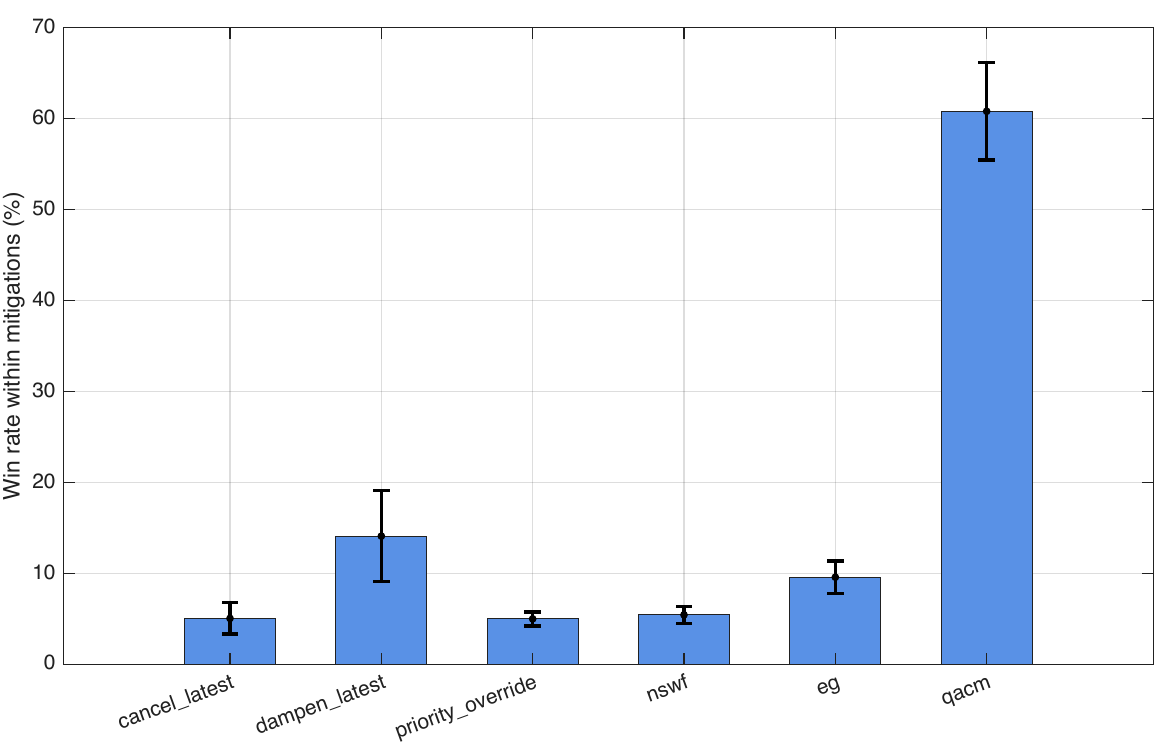}
    \vspace{-0.3cm}
    \caption{Mitigation mechanism selection rates.}
    \label{fig:mechanism_winrate}
\end{figure}

Fig.~\ref{fig:mechanism_winrate} shows the mechanism selection rates. \texttt{qacm} is the dominant mechanism at 60.8\,\% of all selections (CI [55.5,\,66.2]), followed by \texttt{dampen\_latest} at 14.1\,\% (CI [9.1,\,19.1]) and \texttt{eg} at 9.6\,\% (CI [7.8,\,11.4]). \texttt{qacm} dominates because the episodes that clear the $\tau_h$ threshold are predominantly direct conflicts, for which \texttt{qacm} carries the highest efficacy ($\eta=0.88$) combined with the lowest stability and opportunity cost coefficients. The unilateral \texttt{cancel} and \texttt{dampen} mechanisms carry higher stability and opportunity costs, making them inferior in net utility for most direct-conflict episodes under the calibrated model.

The game-theoretic satisfaction rate is high at 93.3\,\% (CI [91.2,\,95.3]), confirming that the compromise-oriented mechanisms produce outcomes aligned with the intended fairness objective.

\textbf{Comparison with baseline strategies.}
Table~\ref{tab:baseline_comparison} compares the proposed framework retroactively against three alternative decision rules applied to the same pre-computed episode utility scores from the 30 runs. \emph{Always-tolerate} never intervenes and achieves zero per-episode utility. \emph{Always-mitigate} applies the best mechanism to every episode: it achieves slightly higher per-episode utility (0.034) but requires $5.1\times$ more interventions (118.3 vs 23.4 per run). \emph{Positive-utility rule} mitigates whenever $U^*(e)>0$: it achieves the highest per-episode utility (0.047) but ignores uncertainty, acting on 44.8\,\% of episodes. The proposed framework extracts 0.132 utility per intervention, $3.9\times$ more than always-mitigate, by confining interventions to episodes that clearly exceed $\tau_h$ with low uncertainty. The 0.020 per-episode gap relative to the positive-utility rule reflects the combined effects of the utility threshold, uncertainty gate, and bounded deferral; this comparison does not isolate their individual contributions.

Both utility columns use the same $U(e,m)$: per-episode utility divides the sum of intervention utilities by all detected episodes, while per-intervention utility divides it by interventions only. Non-intervention contributes zero incremental mitigation utility, not zero unmanaged loss. Table entries are run-level averages, so multiplying averaged rates and utilities need not reproduce the averaged per-episode score. Selectivity improves utility per intervention, while the positive-utility rule has higher utility per episode. No paired significance test is reported, and these model-based comparisons do not establish superiority over independently implemented PACIFISTA, QACM, or learning-based controllers.

\begin{table}[t]
\centering
\caption{Retroactive baseline comparison}
\label{tab:baseline_comparison}
\setlength{\tabcolsep}{4pt}
\begin{tabular}{lccc}
\hline
Policy & Per-Ep utility & Per-Int utility & $N_{\text{int}}$ \\
\hline
Always-tolerate       & 0.000 & ---   & 0     \\
Proposed framework    & 0.027 & 0.132 & 23.4  \\
Always-mitigate       & 0.034 & 0.034 & 118.3 \\
Positive-utility rule & 0.047 & 0.103 & 53.0  \\
\hline
\end{tabular}
\end{table}

\subsection{Interpretability of the Triage Decisions}

\begin{figure}[t]
    \centering
    \includegraphics[width=\columnwidth]{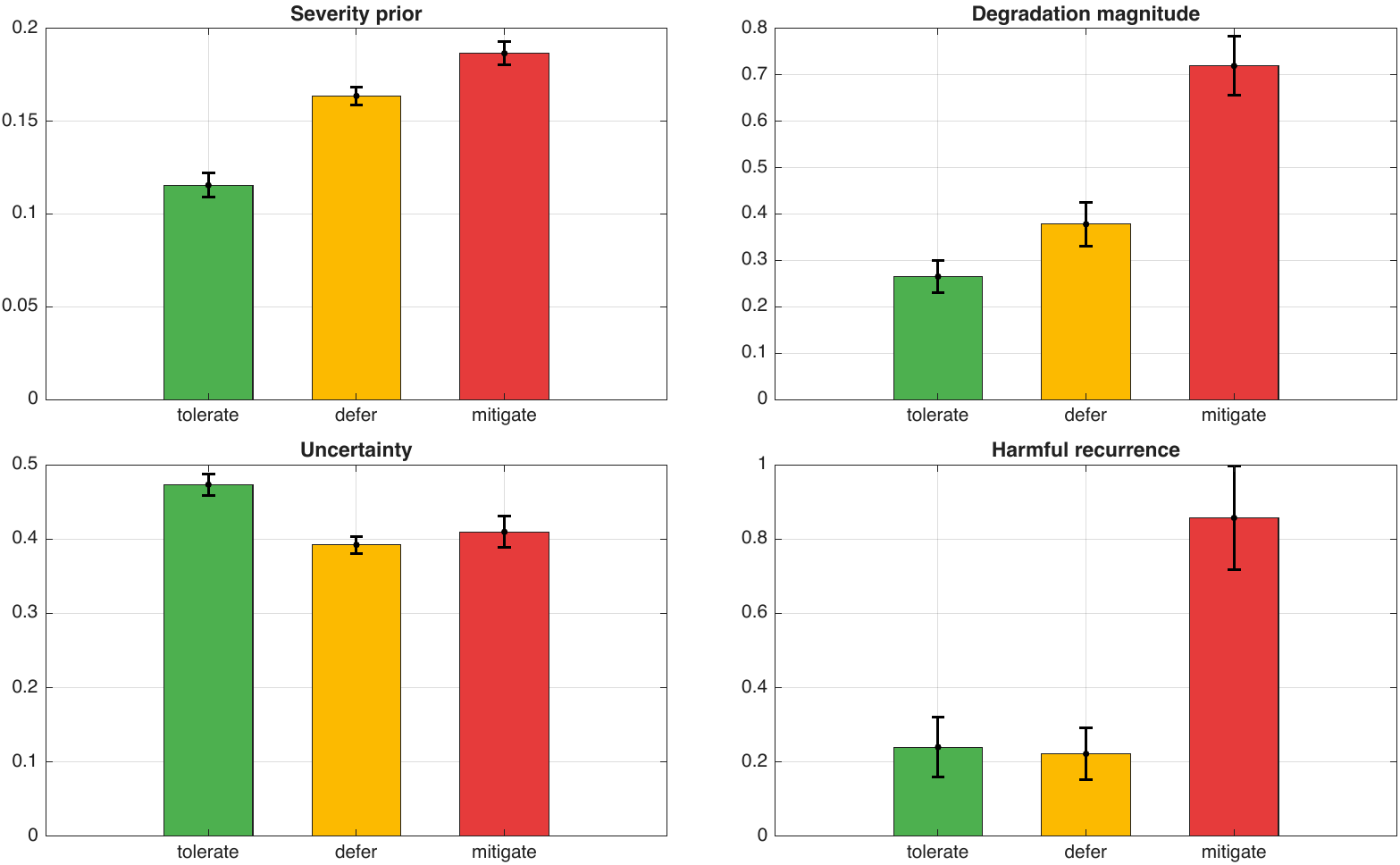}
    \vspace{-0.5cm}
    \caption{Action-conditioned feature distributions.}
    \label{fig:feature_summary}
\end{figure}

The third result is that the triage outcomes are interpretable. Fig.~\ref{fig:feature_summary} shows clear differences between the three action groups. Episodes that are finally mitigated have the highest average severity prior, the largest degradation magnitude, and the strongest harmful recurrence. In contrast, tolerated episodes are associated with lower degradation and weaker harmful recurrence, while also showing the highest uncertainty.

These trends match the intended logic, and align with the design goal of limiting interventions to episodes where the cost is clearly justified. Critically, the deferred group sits between the other two on every feature: higher severity and degradation than tolerated, but lower certainty than mitigated. This ordering is consistent with the decision criteria; it does not independently validate the benefit of deferral.

\subsection{Discussion}

The triage setting is conservative: \texttt{defer} dominates at 66.8\,\%. Unlike ignoring conflicts outright, deferred episodes remain under active observation, and bounded deferral guarantees escalation once $D_{\max}=2$ is reached. This mechanism is consequential: 47.6\,\% (CI [38.3,\,57.0]) of all mitigations across the 30 runs were forced escalations rather than utility-triggered ($U^*(e)>\tau_h$). Forced mitigations carry lower mean net utility ($+0.042$, CI [0.020,\,0.064]) than utility-triggered ones ($+0.210$, CI [0.194,\,0.227]), showing that $D_{\max}$ trades some efficiency for a guarantee that persistent harmful conflicts are not indefinitely postponed by uncertainty; operators may raise $D_{\max}$ for fewer forced interventions or lower it for promptness. These escalation statistics describe the safety--efficiency trade-off under the calibrated model. A two-state ablation and closed-loop re-observation are needed to determine whether deferral reduces unnecessary interventions or improves long-term network stability.

\textbf{Parameter dependence and limitations.}
The thresholds and coefficients encode operator risk preferences. For fixed episode scores, increasing $\tau_h$ or decreasing $u_h$ can only shrink the immediate-mitigation set before forced escalation. Coefficient sensitivity follows directly from $U(e,m)=\eta_{m,c_e}L_{\mathrm{conf}}(e)-\xi_m u_e-C_{\mathrm{mit}}(e,m)$: with other terms fixed, a loss perturbation $\Delta L$ changes a candidate's utility by $\eta_{m,c_e}\Delta L$, while an added cost $\Delta C$ lowers it by $\Delta C$. Decisions near either threshold or near a tie between mechanisms are therefore most sensitive. These are analytical dependencies; action-rate and utility robustness under parameter perturbations have not been measured.

A stale severity prior or noisy degradation measurement can cause false tolerance or unnecessary mitigation. $D_{\max}$ bounds repeated deferral but does not correct false tolerance or guarantee positive utility for forced interventions. Evaluation covers four \acp{xApp}, one topology, and fixed traffic/mobility settings; transfer to other vendors, conflict types, or loads remains untested. The utility-based results assess the scoring policy's internal consistency. Independent throughput, latency, energy, and signalling measurements after executing decisions are needed for closed-loop validation.

\section{Conclusion}
\label{sec:conclusion}

This paper presented a cost-aware conflict triage framework that fills a gap in Open \ac{RAN} conflict management: deciding whether intervention is worth its operational cost, not just how to intervene. By combining unmanaged loss estimation, per-mechanism cost scoring, and bounded deferral, the framework reserves mitigation for episodes with high estimated benefit relative to cost, deferring borderline or uncertain cases instead of acting prematurely. Over 30 Monte Carlo trials with four concurrent \acp{xApp}, this yields internally consistent, selective decisions: 0.132 net utility per intervention ($3.9\times$ an always-mitigate policy) from 23.4 mean interventions per run, with \texttt{qacm} winning 60.8\,\% of selections and bounded deferral ($D_{\max}=2$) accounting for 47.6\,\% of mitigations as a safety net under uncertainty. Future work should add closed-loop \ac{KPI} re-simulation, threshold sensitivity analysis, and validation on a live O-RAN testbed.

\bibliographystyle{IEEEtran}
\bibliography{references}

\end{document}